\documentclass[12pt]{article}
\usepackage{heck}
 
\usepackage{amsmath, amssymb}
\usepackage{bm}
\usepackage{graphicx}
\usepackage{CJK}
\usepackage
{hyperref}
\usepackage[utf8]{inputenc}
\usepackage{cite}
\usepackage{bm}

\graphicspath{{./Figures/}}

\newcommand{\bS}{\ensuremath{\mathbb{S}}}
\newcommand{\bW}{\ensuremath{\mathbb{W}}}
\newcommand{\scN}{\ensuremath{\mathcal{N}}}

\newcommand{\matWs}[5]{\ensuremath{
\mathcal{R}(#1)
\left[\begin{array}{cc}
#5 & #4 \\
#2 & #3 \\
\end{array}
\right]
}}

\begin{document}
\begin{CJK*}{UTF8}{min}

\title{Cluster-Enriched Yang-Baxter Equation \\ from SUSY Gauge Theories}

\authors{\centerline{Masahito Yamazaki (山崎雅人)}}

\date{November, 2016}

\institution{IPMU}{\centerline{Kavli IPMU (WPI), University of Tokyo, Kashiwa, Chiba 277-8583, Japan}}
\institution{Harvard}{\centerline{Center for the Fundamental Laws of Nature,
 Harvard University, Cambridge, MA 02138, USA}}

\abstract{
We propose a new generalization of the Yang-Baxter equation, where the R-matrix depends on cluster $y$-variables in addition to the spectral parameters. We point out that we can construct solutions to this new equation from the recently-found correspondence between Yang-Baxter equations and supersymmetric gauge theories. The $S^2$ partition function of a certain 2d $\mathcal{N}=(2,2)$ quiver gauge theory gives an R-matrix, whereas its FI parameters can be identified with the cluster $y$-variables.
}


\maketitle

\end{CJK*}

\section{Introduction}
The interplay between the physics of supersymmetric gauge theories and the integrable models
(defined here as solutions to the Yang-Baxter equation (YBE) with spectral parameters \cite{Yang:1967bm,Baxter:1972hz}) has been a fascinating subject over the past several decades.

Recently, a new version of such an interplay (the Gauge/YBE correspondence) between 
supersymmetric gauge theories and integrable model has been found \cite{Yamazaki:2012cp,Terashima:2012cx,Yamazaki:2013nra} (see also \cite{Bazhanov:2010kz,Spiridonov:2010em,Bazhanov:2011mz,Xie:2012mr,Yagi:2015lha,Yamazaki:2015voa,Gahramanov:2015cva,Kels:2015bda,Gahramanov:2016ilb}). The correspondence states a rather surprising equivalence between the statistical partition function of a 
classical two-dimensional lattice model on the one hand,
and a supersymmetric partition function of supersymmetric quiver gauge theories, on the other.
The basic idea behind this correspondence is that the Yang-Baxter equation is promoted to a duality (Yang-Baxter duality)
 between supersymmetric gauge theories: since the two theories are dual, their partition functions are the same, and resulting mathematical equality turns out to have an interpretation as YBE. This is arguably one of the most direct correspondence between 
integrable models and supersymmetric gauge theories in the literature.

What is remarkable about this correspondence is that the insights from supersymmetric gauge theories
helps us to find {\it new} integrable models hitherto unknown in the literature.
Indeed, in \cite{Yamazaki:2013nra} a new class of integrable models has been found 
from the lens space ($S^1\times S^3/\mathbb{Z}_r$) index \cite{Benini:2011nc,Razamat:2013opa} of $SU(N)$ quiver gauge theories.
This model is labeled by two integers $N>1$ and $r>0$, has spectral parameters,
and depends on two elliptic parameters $p,q$.\footnote{For $r=1$, this newly-found solution reproduces the solutions found recently in \cite{Bazhanov:2011mz} (see also \cite{Bazhanov:2010kz,Spiridonov:2010em,Bazhanov:2013bh}). For $N=2$ and general $r$, the recent paper \cite{Kels:2015bda} mathematically proves 
integrablity (star-triangle relation) of the model.}
Furthermore one can generalize the story by e.g.\ studying 2d $\mathcal{N}=(2,2)$ quiver gauge theories \cite{Yamazaki:2015voa}.
We expect that there are many more solutions to the YBE yet to be found from this approach.

In this short note, we proceed further and point out that Gauge/YBE correspondence is useful to discover {\it new equations} 
generalizing the standard YBE. What is particularly nice about these equations is that
the resulting equation naturally incorporates the mathematical machinery of 
cluster algebras \cite{FominZelevinsky1,FominZelevinsky4}.
For this reason we call our equations the {\it cluster-enriched Yang-Baxter equation}.

In the rest of this paper,  we first quickly summarize the basic idea of the Gauge/YBE correspondence \ref{sec.review}.
We then construct explicit solutions to the cluster-enriched YBE from
2d $\mathcal{N}=(2,2)$ theories (section \ref{sec.2d}).
The final section (section \ref{sec.conclusion}) contains concluding remarks.

\section{Gauge/YBE Correspondence}\label{sec.review}

Let us quickly summarize the basic idea behind the Gauge/YBE correspondence.
For full details, see \cite{Yamazaki:2012cp,Terashima:2012cx,Yamazaki:2013nra,Yamazaki:2015voa}\footnote{See also the forthcoming review \cite{Yamazaki_GaugeYBE_review}.}.
Readers familiar with these references can safely skip this section.

\subsection{Star-Star Duality and R-matrix}

\begin{figure}[htbp]
\centering\includegraphics[scale=0.4]{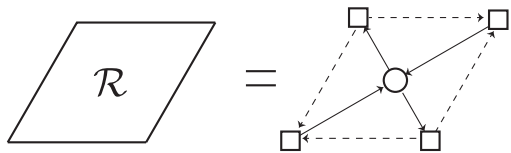}
\caption{The R-matrix is obtained from the partition function of the quiver on the right. Figure reproduced from \cite{Yamazaki:2015voa}.}
\label{fig.R_as_quiver}
\end{figure}

One of the crucial ideas in the Gauge/YBE correspondence is 
to associate a quiver gauge theory (which we call $\mathcal{T}[R]$) to the R-matrix \eqref{fig.R_as_quiver}.
The quiver diagram for the theory $\mathcal{T}[R]$ is shown in Figure \ref{fig.R_as_quiver}.

The quiver of Figure \ref{fig.R_as_quiver} has five nodes. The circle in the middle is the $G$ gauge group,
while the four squares represent the $G$ flavor symmetries; $G$ in this paper is $U(N)$.
This in particular means that the theory $\mathcal{T}[R]$ has 
$G$ flavor symmetries. This is the counterpart of the fact the R-matrix has four indices.
We also note that some of the arrows are dotted---this is meant to be representing a ``half chiral multiplet'' \cite{Yamazaki:2015voa}, namely we take a square root when we discuss partition functions.

\begin{figure}[htbp]
\centering\includegraphics[scale=0.45]{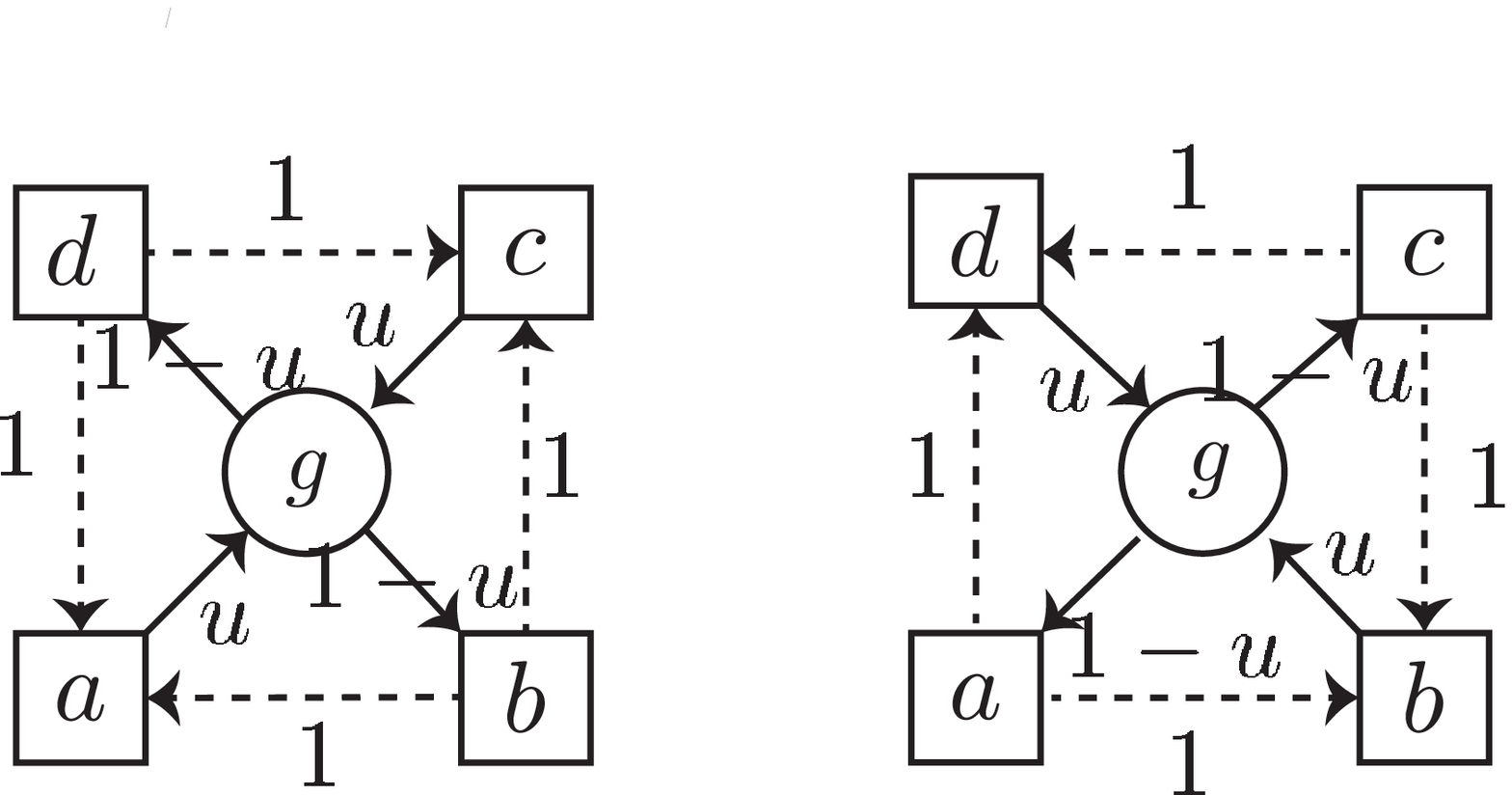}
  \caption{Seiberg-like duality of a quiver gauge theory, which can also be read as the star-star relation of an integrable model. The dotted lines represent the ``half chiral multiplet'', and 
the parameters on the edges (such as $u, 1-u$ and $1$) represent the R-charges of the bifundamental fields. For a closed loop (triangle) their R-charges sum up to two due to superpotential constraints.}
  \label{fig.SeibergD_quiver}
\end{figure}

Now, what is special about the theory $\mathcal{T}[R]$ (and its quiver diagram) is the fact that 
the theory often has dual ({\it star-star dual}), whose graphical representation of the Seiberg(-like) duality
in Figure \ref{fig.SeibergD_quiver}
coincides exactly with that for the star-star relation known from integrable models \cite{Baxter:1986,Bazhanov:1992jqa}.

Whether or not such a duality exists depends crucially on the details of the precise definition of the quiver gauge theory---we have the choice of the 
spacetime dimension, the gauge group at the vertex of the quiver, the number of supersymmetries, etc.
In fact such flexibility is one reason which makes the Gauge/YBE correspondence so rich.

Let us denote the spacetime dimension by $D$.
For $D=4$, we can take $G=SU(N)$ or $U(N)$, and the start-star duality in question is the 
Seiberg duality \cite{Seiberg:1994pq}. For $D=2$, we can also take
$G=U(N)$, where the star-star duality coincides \cite{Yamazaki:2015voa} with the 2d $\scN=(2,2)$ version of the 
Seiberg duality \cite{Benini:2012ui,Gadde:2013dda}.

\subsection{Yang-Baxter Duality and YBE}

In integrable model literature, the star-star equation, applied four times, is known to imply YBE.
We can translate this into supersymmetric gauge theory.
Since the Yang-Baxter equation is about the product of three R-matrices, we can 
glue together three copies of the quiver diagram for $\mathcal{T}[R]$, 
by gauging global symmetries of the theory. The Yang-Baxter duality is then the statement that the
two resulting quiver gauge theories are dual, i.e.\ describe the same physics in the IR fixed point.

Once we obtain the duality, we can compute various supersymmetric partition functions and then obtain solutions to YBE. For $D=4$ we can take $S^1\times S^3/\mathbb{Z}_r$ partition function, for example, and
for $D=2$ the $T^2$ partition function \cite{Gadde:2013dda,Benini:2013xpa}.

The partition function for the theory $\mathcal{T}[R]$ gives the R-matrix\footnote{Here we assumed conditions explicitly listed in \cite{Yamazaki:2015voa}. These conditions are satisfied in the examples in the paper, except as we will see there are some issues in overall normalization.}
\begin{equation}
\begin{split}
\matWs{u}{\bm{a}}{\bm{b}}{\bm{c}}{\bm{d}} 
&:=
\sqrt{
\bW_{1}(\bm{a},\bm{b}) \bW_{1}(\bm{c},\bm{b})
\bW_{1}(\bm{c}, \bm{d}) \bW_{1}(\bm{a}, \bm{d})} \\
&\qquad  \times 
\sqrt{\bS^{\bm{a}} \bS^{\bm{c}}}\, \sum_{\bm{g}}
\bS^{\bm{g}} \, \bW_{u}(\bm{a}, \bm{g}) \bW_{1-u}(\bm{g}, \bm{b}) \bW_{u}(\bm{c}, \bm{g})
\bW_{1-u}(\bm{g}, \bm{d}) \;.
\end{split}
\label{eq.Wdef}
\end{equation}
Here we have chosen the R-charges of the four 
chiral multiplets to be  $u, 1-u, u, 1-u$ in the cyclic order,\footnote{In general we have can 
associate a R-charges $\alpha, \beta, \gamma, \delta$ to the four bifundamental multiplets,
with possibly further constraints in order to make sure that these parametrize global symmetries of the theory.
This defines a more general R-matrix depending on multiple parameters.
See \cite{Yamazaki:2015voa} for more details.
} and $u$ plays the role of the spectral parameter in integrable models.

In the expression \eqref{eq.Wdef},
the factor $\bW_{r_e}(t(e), h(e))$
is the 1-loop determinant for a matter supersymmetric multiplet associated with an edge $e$ with endpoints $t(e)$ and $h(e)$, and with R-charge $r_e$.
Similarly, $\bS^v$ is the classical as well as the 1-loop contribution, for a gauge supersymmetric multiplet associated with a vertex $v$ of the quiver diagram.

The equivalence of the partition functions originating from the Yang-Baxter duality now 
is precisely the Yang-Baxter equation:
\begin{equation}
\begin{split}
& \sum_{\bm{g}} \matWs{u}{\bm{a}}{\bm{b}}{\bm{g}}{\bm{f}}
\matWs{u+v}{\bm{g}}{\bm{b}}{\bm{c}}{\bm{d}}
\matWs{v}{\bm{f}}{\bm{g}}{\bm{d}}{\bm{e}}
\\
& \qquad \qquad=\sum_{\bm{g}}
 \matWs{v}{\bm{a}}{\bm{b}}{\bm{c}}{\bm{g}}
 \matWs{u+v}{\bm{f}}{\bm{a}}{\bm{g}}{\bm{e}}
\matWs{u}{\bm{g}}{\bm{c}}{\bm{d}}{\bm{e}} \ .
\end{split}
\label{eq.YBE}
\end{equation}

\section{Cluster-Enriched YBE from 2d \texorpdfstring{$\mathcal{N}=(2,2)$} {N=(2,2)} Theories}\label{sec.2d}

\subsection{FI Parameters as Cluster Variables}

In this section we consider 2d $\mathcal{N}=(2,2)$ quiver gauge theory,
where the gauge group at the vertex of the quiver diagram is $U(N)$, with the same value of $N$ 
for all vertices.\footnote{One advantage of this choice is that the rank of the gauge group,
and hence number of the components of the spin of the integrable model at a lattice site, 
is preserved by the Seiberg-like duality. It is straightfoward to allow different gauge groups to different nodes. Such a generalization will further refine our YBE. Note that in \cite{Benini:2014mia}
the ranks of the gauge groups transform as cluster $x$-variables.}

The star-star duality in this case will then be
the 2d version of the Seiberg duality, 
as proposed in \cite{Benini:2012ui,Gadde:2013dda} (see also  \cite{Benini:2013xpa,Jia:2014ffa}), 
and the integrable model associated with the $T^2$ partition function \cite{Gadde:2013dda,Benini:2013nda,Benini:2013xpa} was studied in \cite{Yamazaki:2015voa}. Here we instead study their $S^2$ partition functions \cite{Benini:2012ui,Doroud:2012xw}.

Compared with the $T^2$ partition function, 
the $S^2$ partition function depends on a set of  extra parameters, the complexified FI parameters,
which transform non-trivially under the Seiberg-duality, and hence under the Yang-Baxter duality.

Recall that in 2d $\mathcal{N}=(2,2)$ theories the FI (Fayet-Iliopoulos) parameter $r$ is naturally complexified
by the theta angle $\theta$,\footnote{
The relevant term in the Lagrangian is given by
\begin{align}
\mathcal{L}_{\rm FI}
= -r D + \theta F_{01} \;,
\label{eq.complexFI}
\end{align}
where $D$ is an auxiliary field in the $\scN=(2,2)$ vector multiplet.
} into an exponentiated variable $y$:\footnote{This was denoted as $z$ in \cite{Benini:2014mia,Yamazaki:2015voa}.}
\begin{align}
y:=(-1)^N e^{r+i \theta} \;.
\end{align}
The factor of $(-1)^N$ is inserted for a better match with cluster algebra literature.

Now the highly nontrivial statement found in \cite{Benini:2014mia} was that
the complexfied FI parameters for the Seiberg-like dual pair theories (Figure \ref{fig.SeibergD_quiver}) are given by
\begin{align}
\begin{split}
&y'_g= z_g^{-1} \;, \\
&y'_a= z_a(1+z_g) \;, \quad
y'_b= z_b(1+z_g^{-1})^{-1} \;, \quad
y'_c= z_c(1+z_g) \;, \quad
y'_d= z_d(1+z_g^{-1})^{-1} \;, \quad
\label{z_inverse}
\end{split}
\end{align}
with primed (unprimed) variables representing the parameters for the after (before) the 
Seiberg-like duality. Interestingly, 
 that this transformation rule is the same 
as the transformation rules of the ``cluster $y$-variable'' under a ``mutation'' of the quiver.

\begin{figure}[htbp]
\centering\includegraphics[scale=0.4]{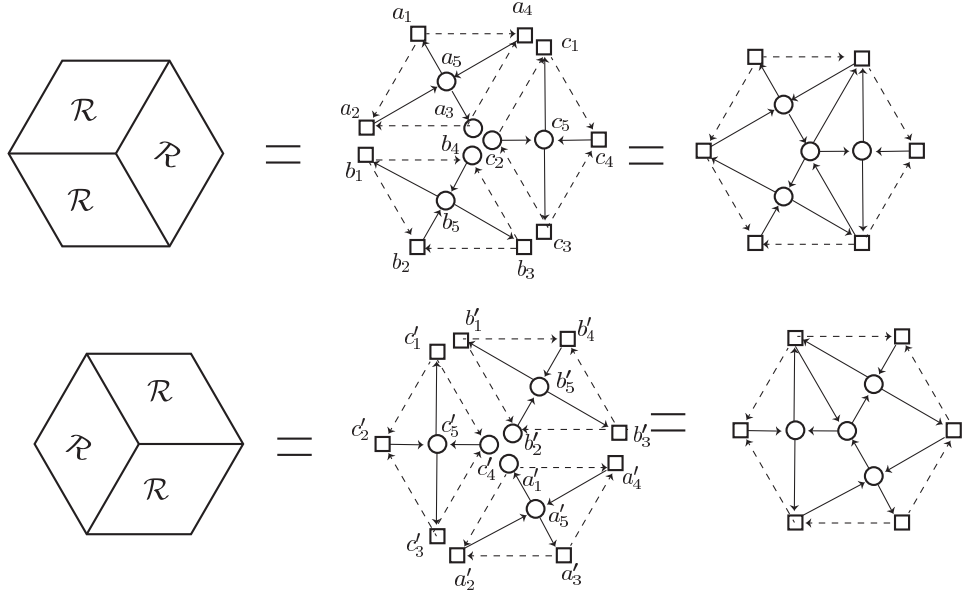}
\caption{Labeling of the complexified FI parameters. Each R-matrix has five vertices,
and hence five FI parameters, which we label as $\vec{a}=(a_1, \ldots, a_5)$, $\vec{b}$ and $\vec{c}$. On the other side of the YBE we use the similar labeling, with the primed variables.}
\label{fig.R_glue_abc}
\end{figure}

By using this transformation property we can easily compute the transformation 
properties of the complexfied $y$-variables.
Let us parametrize the exponentiated complexified FI parameters 
as in Figure \ref{fig.R_glue_abc}. Then the $y$-variables at the $10$ vertices are given by
\begin{align}
\begin{split}
y_1 &\to a_3 b_4 c_2\;,\quad
y_2 \to a_5\;,\quad
y_3 \to b_5 \;,\quad
y_4 \to c_5 \;,\quad
y_5 \to a_1 \;,\\
y_6 &\to a_2 b_1\;,\quad
y_7 \to b_2\;,\quad
y_8 \to b_3 c_3\;,\quad
y_9 \to c_4\;,\quad
y_{10} \to c_1 a_4\;,
\end{split}
\end{align}      
and
\begin{align}
\begin{split}
y_1' &\to a_5' \;,\quad
y_2' \to a_1' b_2' c_4' \;,\quad
y_3' \to c_5' \;,\quad
y_4' \to b_5' \;,\quad
y_5' \to b_1' c_1' \;,\\
y_6' &\to c_2'\;,\quad
y_7' \to a_2' c_3'\;,\quad
y_8' \to a_3' \;,\quad
y_9' \to a_4' b_3'\;,\quad
y_{10}' \to b_4'\;,
\end{split}
\end{align} 
where we used that fact that we need to combine the FI parameters.
when we glue quivers (we simply add terms in the Lagrangian). 

\begin{figure}[htbp]
\centering\includegraphics[scale=0.4]{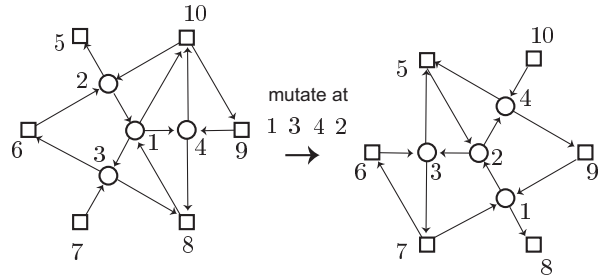}
\caption{The labeling of the vertices, on both sides of the YBE. We associate a complexified FI parameter $y_i$ ($y_i'$) to the $i$-th vertex of the quiver on the left (right).}
\label{fig.R_glue_label}
\end{figure}

After 4 mutations at vertices $1,3,4,2$ (in this order; see Figure \ref{fig.R_glue_label}), the $y$-variables transform as
\begin{align}
\begin{split}
y_1'&=\frac{y_1 y_2 y_3
   y_4}{y_1^2 y_2 (y_3+1)
   (y_4+1)+y_1 (y_2
   (y_3+y_4+2)+1)+y_2+1}\;, \\
y_2&'=\frac{y_1+
   1}{y_2 (y_1 y_3+y_1+1)
   (y_1 y_4+y_1+1)}\;,\\
y_3&'=\frac{y_1^2
   y_2 (y_3+1) (y_4+1)+y_1
   (y_2
   (y_3+y_4+2)+1)+y_2+1}{y_1
   y_3}\;,\\
y_4&'=\frac{y_1^2 y_2 (y_3+1)
   (y_4+1)+y_1 (y_2
   (y_3+y_4+2)+1)+y_2+1}{y_1
   y_4}\;,\\
y_5&'=   \frac{y_2 y_5 (y_1
   y_3+y_1+1) (y_1
   y_4+y_1+1)}{y_1^2 y_2
   (y_3+1) (y_4+1)+y_1 (y_2
   (y_3+y_4+2)+1)+y_2+1}\;,\\
y_6&'=   \frac{y_1
   y_3 y_6}{y_1
   y_3+y_1+1}\;,\\
y_7&'=   \frac{y_1 y_3
   y_7}{y_1+1}+y_7\;,
   \\
y_8&'=   (y_1+1)
   y_8\;,\\
y_9&'=   \frac{y_1 y_4
   y_9}{y_1+1}+y_9\;,\\
y_{10}&'=   \frac{y_1
   y_{10} y_4}{y_1
   y_4+y_1+1}\;.
\label{y_transf}   
\end{split}
\end{align}

The R-matrix now depends explicitly on the FI parameters, 
\begin{align}
\matWs{u; \vec{a}}{\bm{a}}{\bm{b}}{\bm{c}}{\bm{d}} \;,
\end{align}
where $\vec{a}=(a_1, a_2, \ldots, a_5)$ is a set of the FI parameters for the five vertices of 
the theory $\mathcal{T}[R]$, and primed/unprimed variables should satisfy the constraint
\eqref{y_transf}.

The identity representing the Yang-Baxter duality reads
\begin{equation}
\begin{split}
& \sum_{\bm{g}} \matWs{u; \vec{a}}{\bm{a}}{\bm{b}}{\bm{g}}{\bm{f}}
\matWs{u+v; \vec{b}}{\bm{g}}{\bm{b}}{\bm{c}}{\bm{d}}
\matWs{v; \vec{c}}{\bm{f}}{\bm{g}}{\bm{d}}{\bm{e}}
\\
& \qquad \qquad=\sum_{\bm{g}}
 \matWs{v; \vec{c}'}{\bm{a}}{\bm{b}}{\bm{c}}{\bm{g}}
 \matWs{u+v; \vec{b}'}{\bm{f}}{\bm{a}}{\bm{g}}{\bm{e}}
\matWs{u; \vec{a}'}{\bm{g}}{\bm{c}}{\bm{d}}{\bm{e}} \ .
\end{split}
\label{eq.YBE_cluster}
\end{equation}

There is one subtlety in \eqref{eq.YBE_cluster}. The $S^2$ partition function has an ambiguity of the K\"{a}hler transformation \cite{Gomis:2015yaa}:
\begin{align}
\log Z_{S^2}\to \log Z_{S^2}+f(y)+\bar{f}(\bar{y}) \;,
\end{align} 
where $f(y)$ ($\bar{f}(\bar{y})$) is a holomorphic (anti-holomorphic) function of $y$.
This means that more naturally the identity \eqref{eq.YBE_cluster}
should be interpreted as an identity up to this ambiguity.

If we wish we can eliminate this ambiguity by using the explicit 
numerical factor derived in \cite{Benini:2014mia}. The result is that 
\eqref{eq.YBE_cluster} holds including the overall factor 
if we further impose the condition
\begin{align}
f(y_1)\, f\left(\frac{y_1 y_3}{1+y_1}\right) \, f\left(\frac{y_1 y_4}{1+y_1}\right) \,
f\left(\frac{y_2(1+y_1 +y_1 y_3)(1+y_1 +y_1 y_4)}{1+y_1}\right)=1
\label{Kahler_cond}
 \;,
\end{align}
where $f(y)$ is a function denoted by $f^{(r)}_{\rm ctc}$ in \cite{Benini:2014mia}.
It would be nice to better understand the cluster-algebraic significance of this constraint.

\subsection{Expression for R-matrix}

Let us also write down explicit expression for the R-matrix.

The $S^2$ partition function is represented as an integral over the 
Cartan of the gauge group, which for a $U(N)$ gauge group
is parametrized by $N$ parameters.
We in addition have a 
monopole flux, a set of $N$ integers, and we take a sum over them.
Correspondingly, the integrable model has spins $s_v$
taking values in $\mathbb{R}^N \times \mathbb{Z}^N$.
First, we have $N$ continuous variables $\sigma_{v,i}$,
corresponding to the values of the Coulomb branch scalar
inside the $\mathcal{N}=(2,2)$ vector multiplet.
We also have $N$ discrete  variables $m_{v,i}$,
corresponding to magnetic fluxes on $S^2$. Correspondingly, the sum over 
$s_v$ reads
\begin{align}
\sum_v \to \sum_{m_{v,i}} \int \prod_i d\sigma_{v,i} \;.
\end{align}

For an edge $e$ corresponding to a 2d $\scN=(2,2)$ chiral multiplet,
the Boltzmann weights is given by
\begin{align}
\bW_e(s_{t(e)}, s_{h(e)})= \prod_{i,j}\frac{
\Gamma\left( \frac{r_e}{2} -  i \sigma_{t(e),i, j}-m_{t(e),i, j}
\right)
}
{
\Gamma\left( 1-\frac{r_e}{2} +  i \sigma_{t(e),i, j}-m_{t(e),i, j}
\right) 
} \;,
\end{align}
where $\sigma_{v,i, j}=:\sigma_{v,i}-\sigma_{v,j}$
and $m_{v,i, j}=:m_{v,i}-m_{v,j}$.

For a vertex $v$ corresponding to a 2d $\scN=(2,2)$ vector multiplet, we have
\begin{align}
\bS^v(s_v)=\bS_{\rm gauge}^v(s_v) \,\, \bS_{\rm FI}^v(s_v)
\end{align}
with\footnote{
In $\bS^{\rm FI}_v(s_v)$ we included a sign factor $(-1)^{(N-1) \sum_j m_{v, j}}$ as suggested by \cite{Dimofte:2011py}, for better comparsion with \cite{Benini:2014mia}. For our considerations only the identity of the partition functions for the star-star (and Yang-Baxter) duals is of importance, 
and this is not affected by such a sign.}
\begin{align}
&\bS_{\rm gauge}^v(s_v):=\frac{1}{N!} \prod_{i\neq j} \left[\left(\sigma_{v,i}-\sigma_{v,j}\right)^2+\left( \frac{m_i-m_j}{2}\right)^2 \right]
 \;, \\
&\bS_{\rm FI}^v(s_v)
:=(-1)^{(N-1) \sum_j m_{v, j}}e^{2i \left(\sum_j \sigma_{v,j}\right) t +i\theta \left(\sum_j m_{v,j}\right) } \;.
\end{align}

\section{Discussion}\label{sec.conclusion}

In this paper, we provided solutions to a version the Yang-Baxter equation
where the R-matrix also depends on a cluster variable (or its tropical counterpart).

There could be other examples of supersymmetric gauge theories leading to novel cluster-algebra-enrichment of YBE. For example, we can appeal to the Giveon-Kutasov duality \cite{Giveon:2008zn} for 3d $\mathcal{N}=2$ Chern-Simons-matter theories. In this case,
the ranks of the gauge groups (the number of components of spins) and the Chern-Simons levels (extra parameter at a vertex of the quiver) transform as tropical $x$- and $y$-variables, respectively  \cite{Xie:2013lya}, and for example  the $S^1\times S^2$ partition function \cite{Kim:2009wb,Imamura:2011su} of the theory gives some refinement of the YBE.
It is also a natural question if there is any connection with another cluster algebra structure found in the literature, 
namely ``3d cluster $\mathcal{N}=2$ theories'' of \cite{Terashima:2013fg,Gang:2015wya},
where a 3d $\mathcal{N}=2$ theories was associated with a mutation sequence of a quiver.

As we have seen, the cluster-enriched YBE is a natural equation 
from the viewpoint of supersymmetric gauge theory.
The real significance of the equation, however, is not clear as of this writing, since for example
the equation does not ensure existence of an infinite number of conserved charges.
In this respect one useful analogy is another generalization of the YBE, the so-called
dynamical YBE. Historically, the dynamical YBE 
first appeared in 1983 in the study of Liouville theory \cite{Gervais:1983ry}.
However, it was only 10 years later when people begin to appreciate the underlying mathematical structure of the 
dynamical YBEs \cite{Felder:1994pb,Felder:1994be}.
While or not whether the history repeats itself only time will tell, it is fair to say that one should take this lesson 
seriously.

\section*{Acknowledgements}
The author would like to thank Wenbin Yan for collaboration in \cite{Yamazaki:2015voa}, which lead the author to the current project. He would also like to thank the organizers and the audience
of the workshop ``Integrability in Gauge and String Theory 2015'' (Imperial college), where the results of this paper was announced during the author's talk.
This research is supported in part by the WPI Research Center
Initiative (MEXT, Japan), by JSPS Program for Advancing Strategic
International Networks to Accelerate the Circulation of Talented Researchers,
by JSPS KAKENHI (15K17634), by JSPS-NRF Joint Research Project.

\bibliographystyle{nb}
\bibliography{clusterYBE}

\def\cprime{$'$}
\begin{thebibliography}{10}
\providecommand{\href}[2]{#2}
\providecommand{\arxivref}[2]{\href{http://arxiv.org/abs/#1}{#2}}
\providecommand{\doiref}[2]{\href{http://dx.doi.org/#1}{#2}}
\providecommand{\nbbstauthor}[1]{#1}
\providecommand{\nbbstjournal}[1]{\textsf{#1}}
\providecommand{\nbbsttitle}[1]{\textit{#1}}
\providecommand{\nbbsturl}[1]{\texttt{#1}}
\providecommand{\nbbsteprint}[1]{\texttt{#1}}
\providecommand{\nbbststyle}{\raggedright\small\parskip0pt}
\nbbststyle

\bibitem{Yang:1967bm}
\nbbstauthor{C.-N.~Yang},
\nbbsttitle{``{Some exact results for the many body problems in one dimension
  with repulsive delta function interaction}''},
\nbbstjournal{\doiref{10.1103/PhysRevLett.19.1312}{Phys.~Rev.~Lett.~19,~1312~(1967)}}.

\bibitem{Baxter:1972hz}
\nbbstauthor{R.~J.~Baxter},
\nbbsttitle{``{Partition function of the eight vertex lattice model}''},
\nbbstjournal{\doiref{10.1016/0003-4916(72)90335-1}{Annals~Phys.~70,~193~(1972)}},
[Annals Phys.281,187(2000)].

\bibitem{Yamazaki:2012cp}
\nbbstauthor{M.~Yamazaki},
\nbbsttitle{``{Quivers, YBE and 3-manifolds}''},
\nbbstjournal{\doiref{10.1007/JHEP05(2012)147}{JHEP~1205,~147~(2012)}},
\nbbsteprint{\arxivref{1203.5784}{arxiv:1203.5784}}.

\bibitem{Terashima:2012cx}
\nbbstauthor{Y.~Terashima and M.~Yamazaki},
\nbbsttitle{``{Emergent 3-manifolds from 4d Superconformal Indices}''},
\nbbstjournal{\doiref{10.1103/PhysRevLett.109.091602}{Phys.~Rev.~Lett.~109,~091602~(2012)}},
\nbbsteprint{\arxivref{1203.5792}{arxiv:1203.5792}}.

\bibitem{Yamazaki:2013nra}
\nbbstauthor{M.~Yamazaki},
\nbbsttitle{``{New Integrable Models from the Gauge/YBE Correspondence}''},
\nbbstjournal{\doiref{10.1007/s10955-013-0884-8}{J.~Statist.~Phys.~154,~895~(2014)}},
\nbbsteprint{\arxivref{1307.1128}{arxiv:1307.1128}}.

\bibitem{Bazhanov:2010kz}
\nbbstauthor{V.~V.~Bazhanov and S.~M.~Sergeev},
\nbbsttitle{``{A Master solution of the quantum Yang-Baxter equation and
  classical discrete integrable equations}''},
\nbbsteprint{\arxivref{1006.0651}{arxiv:1006.0651}}.

\bibitem{Spiridonov:2010em}
\nbbstauthor{V.~Spiridonov},
\nbbsttitle{``{Elliptic beta integrals and solvable models of statistical
  mechanics}''},
\nbbsteprint{\arxivref{1011.3798}{arxiv:1011.3798}}.

\bibitem{Bazhanov:2011mz}
\nbbstauthor{V.~V.~Bazhanov and S.~M.~Sergeev},
\nbbsttitle{``{Elliptic gamma-function and multi-spin solutions of the
  Yang-Baxter equation}''},
\nbbstjournal{\doiref{10.1016/j.nuclphysb.2011.10.032}{Nucl.~Phys.~B856,~475~(2012)}},
\nbbsteprint{\arxivref{1106.5874}{arxiv:1106.5874}}.

\bibitem{Xie:2012mr}
\nbbstauthor{D.~Xie and M.~Yamazaki},
\nbbsttitle{``{Network and Seiberg Duality}''},
\nbbstjournal{\doiref{10.1007/JHEP09(2012)036}{JHEP~1209,~036~(2012)}},
\nbbsteprint{\arxivref{1207.0811}{arxiv:1207.0811}}.

\bibitem{Yagi:2015lha}
\nbbstauthor{J.~Yagi},
\nbbsttitle{``{Quiver gauge theories and integrable lattice models}''},
\nbbsteprint{\arxivref{1504.04055}{arxiv:1504.04055}}.

\bibitem{Yamazaki:2015voa}
\nbbstauthor{M.~Yamazaki and W.~Yan},
\nbbsttitle{``{Integrability from 2d ${\mathcal{N}}=(2,2)$ dualities}''},
\nbbstjournal{\doiref{10.1088/1751-8113/48/39/394001}{J.~Phys.~A48,~394001~(2015)}},
\nbbsteprint{\arxivref{1504.05540}{arxiv:1504.05540}}.

\bibitem{Gahramanov:2015cva}
\nbbstauthor{I.~Gahramanov and V.~P.~Spiridonov},
\nbbsttitle{``{The star-triangle relation and 3d superconformal indices}''},
\nbbstjournal{\doiref{10.1007/JHEP08(2015)040}{JHEP~1508,~040~(2015)}},
\nbbsteprint{\arxivref{1505.00765}{arxiv:1505.00765}}.

\bibitem{Kels:2015bda}
\nbbstauthor{A.~P.~Kels},
\nbbsttitle{``{New solutions of the star-triangle relation with discrete and
  continuous spin variables}''},
\nbbstjournal{\doiref{10.1088/1751-8113/48/43/435201}{J.~Phys.~A48,~435201~(2015)}},
\nbbsteprint{\arxivref{1504.07074}{arxiv:1504.07074}}.

\bibitem{Gahramanov:2016ilb}
\nbbstauthor{I.~Gahramanov and A.~P.~Kels},
\nbbsttitle{``{The star-triangle relation, lens partition function, and
  hypergeometric sum/integrals}''},
\nbbsteprint{\arxivref{1610.09229}{arxiv:1610.09229}}.

\bibitem{Benini:2011nc}
\nbbstauthor{F.~Benini, T.~Nishioka and M.~Yamazaki},
\nbbsttitle{``{4d Index to 3d Index and 2d TQFT}''},
\nbbstjournal{\doiref{10.1103/PhysRevD.86.065015}{Phys.~Rev.~D86,~065015~(2012)}},
\nbbsteprint{\arxivref{1109.0283}{arxiv:1109.0283}}.

\bibitem{Razamat:2013opa}
\nbbstauthor{S.~S.~Razamat and B.~Willett},
\nbbsttitle{``{Global Properties of Supersymmetric Theories and the Lens
  Space}''},
\nbbstjournal{\doiref{10.1007/s00220-014-2111-0}{Commun.~Math.~Phys.~334,~661~(2015)}},
\nbbsteprint{\arxivref{1307.4381}{arxiv:1307.4381}}.

\bibitem{Bazhanov:2013bh}
\nbbstauthor{V.~V.~Bazhanov, A.~P.~Kels and S.~M.~Sergeev},
\nbbsttitle{``{Comment on star-star relations in statistical mechanics and
  elliptic gamma-function identities}''},
\nbbsteprint{\arxivref{1301.5775}{arxiv:1301.5775}}.

\bibitem{FominZelevinsky1}
\nbbstauthor{S.~Fomin and A.~Zelevinsky},
\nbbsttitle{``Cluster algebras {I}: {Foundations}''},
\nbbstjournal{J.~Amer.~Math.~Soc.~15,~497~(2002)}.

\bibitem{FominZelevinsky4}
\nbbstauthor{S.~Fomin and A.~Zelevinsky},
\nbbsttitle{``Cluster algebras. {IV}. {C}oefficients''},
\nbbstjournal{\doiref{10.1112/S0010437X06002521}{Compos.~Math.~143,~112~(2007)}}.

\bibitem{Yamazaki_GaugeYBE_review}
\nbbstauthor{M.~Yamazaki},
\nbbsttitle{``{to appear}''}.

\bibitem{Baxter:1986}
\nbbstauthor{R.~Baxter},
\nbbsttitle{``{The Yang-Baxter Equations and the Zamolodchikov Model}''},
\nbbstjournal{Physica~D18,~321~(1986)}.

\bibitem{Bazhanov:1992jqa}
\nbbstauthor{V.~Bazhanov and R.~Baxter},
\nbbsttitle{``{New solvable lattice models in three-dimensions}''},
\nbbstjournal{\doiref{10.1007/BF01050423}{J.~Statist.~Phys.~69,~453~(1992)}}.

\bibitem{Seiberg:1994pq}
\nbbstauthor{N.~Seiberg},
\nbbsttitle{``{Electric - magnetic duality in supersymmetric nonAbelian gauge
  theories}''},
\nbbstjournal{\doiref{10.1016/0550-3213(94)00023-8}{Nucl.~Phys.~B435,~129~(1995)}},
\nbbsteprint{\arxivref{hep-th/9411149}{hep-th/9411149}}.

\bibitem{Benini:2012ui}
\nbbstauthor{F.~Benini and S.~Cremonesi},
\nbbsttitle{``{Partition functions of $\mathcal{N}=(2,2)$ gauge theories on
  $S^2$ and vortices}''},
\nbbsteprint{\arxivref{1206.2356}{arxiv:1206.2356}}.

\bibitem{Gadde:2013dda}
\nbbstauthor{A.~Gadde and S.~Gukov},
\nbbsttitle{``{2d Index and Surface operators}''},
\nbbstjournal{\doiref{10.1007/JHEP03(2014)080}{JHEP~1403,~080~(2014)}},
\nbbsteprint{\arxivref{1305.0266}{arxiv:1305.0266}}.

\bibitem{Benini:2013xpa}
\nbbstauthor{F.~Benini, R.~Eager, K.~Hori and Y.~Tachikawa},
\nbbsttitle{``{Elliptic genera of 2d N=2 gauge theories}''},
\nbbsteprint{\arxivref{1308.4896}{arxiv:1308.4896}}.

\bibitem{Benini:2014mia}
\nbbstauthor{F.~Benini, D.~S.~Park and P.~Zhao},
\nbbsttitle{``{Cluster algebras from dualities of 2d N=(2,2) quiver gauge
  theories}''},
\nbbsteprint{\arxivref{1406.2699}{arxiv:1406.2699}}.

\bibitem{Jia:2014ffa}
\nbbstauthor{B.~Jia, E.~Sharpe and R.~Wu},
\nbbsttitle{``{Notes on nonabelian (0,2) theories and dualities}''},
\nbbstjournal{\doiref{10.1007/JHEP08(2014)017}{JHEP~1408,~017~(2014)}},
\nbbsteprint{\arxivref{1401.1511}{arxiv:1401.1511}}.

\bibitem{Benini:2013nda}
\nbbstauthor{F.~Benini, R.~Eager, K.~Hori and Y.~Tachikawa},
\nbbsttitle{``{Elliptic genera of two-dimensional N=2 gauge theories with
  rank-one gauge groups}''},
\nbbstjournal{\doiref{10.1007/s11005-013-0673-y}{Lett.~Math.~Phys.~104,~465~(2014)}},
\nbbsteprint{\arxivref{1305.0533}{arxiv:1305.0533}}.

\bibitem{Doroud:2012xw}
\nbbstauthor{N.~Doroud, J.~Gomis, B.~Le~Floch and S.~Lee},
\nbbsttitle{``{Exact Results in D=2 Supersymmetric Gauge Theories}''},
\nbbstjournal{\doiref{10.1007/JHEP05(2013)093}{JHEP~1305,~093~(2013)}},
\nbbsteprint{\arxivref{1206.2606}{arxiv:1206.2606}}.

\bibitem{Gomis:2015yaa}
\nbbstauthor{J.~Gomis, P.-S.~Hsin, Z.~Komargodski, A.~Schwimmer, N.~Seiberg and
  S.~Theisen},
\nbbsttitle{``{Anomalies, Conformal Manifolds, and Spheres}''},
\nbbstjournal{\doiref{10.1007/JHEP03(2016)022}{JHEP~1603,~022~(2016)}},
\nbbsteprint{\arxivref{1509.08511}{arxiv:1509.08511}}.

\bibitem{Dimofte:2011py}
\nbbstauthor{T.~Dimofte, D.~Gaiotto and S.~Gukov},
\nbbsttitle{``{3-Manifolds and 3d Indices}''},
\nbbsteprint{\arxivref{1112.5179}{arxiv:1112.5179}}.

\bibitem{Giveon:2008zn}
\nbbstauthor{A.~Giveon and D.~Kutasov},
\nbbsttitle{``{Seiberg Duality in Chern-Simons Theory}''},
\nbbstjournal{\doiref{10.1016/j.nuclphysb.2008.09.045}{Nucl.~Phys.~B812,~1~(2009)}},
\nbbsteprint{\arxivref{0808.0360}{arxiv:0808.0360}}.

\bibitem{Xie:2013lya}
\nbbstauthor{D.~Xie},
\nbbsttitle{``{Three dimensional Seiberg-like duality and tropical cluster
  algebra}''},
\nbbsteprint{\arxivref{1311.0889}{arxiv:1311.0889}}.

\bibitem{Kim:2009wb}
\nbbstauthor{S.~Kim},
\nbbsttitle{``{The Complete superconformal index for N=6 Chern-Simons
  theory}''},
\nbbstjournal{\doiref{10.1016/j.nuclphysb.2009.06.025}{Nucl.~Phys.~B821,~241~(2009)}},
\nbbsteprint{\arxivref{0903.4172}{arxiv:0903.4172}}.

\bibitem{Imamura:2011su}
\nbbstauthor{Y.~Imamura and S.~Yokoyama},
\nbbsttitle{``{Index for three dimensional superconformal field theories with
  general R-charge assignments}''},
\nbbstjournal{\doiref{10.1007/JHEP04(2011)007}{JHEP~1104,~007~(2011)}},
\nbbsteprint{\arxivref{1101.0557}{arxiv:1101.0557}}.

\bibitem{Terashima:2013fg}
\nbbstauthor{Y.~Terashima and M.~Yamazaki},
\nbbsttitle{``{3d N=2 Theories from Cluster Algebras}''},
\nbbstjournal{\doiref{10.1093/ptep/PTT115}{PTEP~023,~B01~(2014)}},
\nbbsteprint{\arxivref{1301.5902}{arxiv:1301.5902}}.

\bibitem{Gang:2015wya}
\nbbstauthor{D.~Gang, N.~Kim, M.~Romo and M.~Yamazaki},
\nbbsttitle{``{Aspects of Defects in 3d-3d Correspondence}''},
\nbbsteprint{\arxivref{1510.05011}{arxiv:1510.05011}}.

\bibitem{Gervais:1983ry}
\nbbstauthor{J.-L.~Gervais and A.~Neveu},
\nbbsttitle{``{Novel Triangle Relation and Absence of Tachyons in Liouville
  String Field Theory}''},
\nbbstjournal{\doiref{10.1016/0550-3213(84)90469-3}{Nucl.~Phys.~B238,~125~(1984)}}.

\bibitem{Felder:1994pb}
\nbbstauthor{G.~Felder},
\nbbsttitle{``{Conformal field theory and integrable systems associated to
  elliptic curves}''},
\nbbsteprint{\arxivref{hep-th/9407154}{hep-th/9407154}}.

\bibitem{Felder:1994be}
\nbbstauthor{G.~Felder},
\nbbsttitle{``{Elliptic quantum groups}''},
\nbbsteprint{\arxivref{hep-th/9412207}{hep-th/9412207}},
in: \nbbsttitle{``{Mathematical physics. Proceedings, 11th International
  Congress, Paris, France, July 18-22, 1994}''}.

\end{thebibliography}

\end{document}